\documentclass{PoS}

\title{Enhancement of $H\to\gamma\gamma$  from charged Higgs bosons in the Higgs Triplet Model}

\ShortTitle{Enhancement of  $H\to \gamma\gamma$ ...}

\author{{Andrew G. Akeroyd}%
         \thanks{AGA was supported by a MC IIF, FP7-
PEOPLE-2009-IIF, Contract No. 252263.}\\
        School of Physcs \& Astronomy, University of Southampton, Highfield SO17 1BJ, UK\\
        E-mail: \email{A.G.Akeroyd@soton.ac.uk}}

\author{\speaker{Stefano MORETTI}%
         \thanks{ SM is partially supported through the NExT Institute.}\\
        School of Physcs \& Astronomy, University of Southampton, Highfield SO17 1BJ, UK\\
        E-mail: \email{S.Moretti@soton.ac.uk}}

\abstract{We explain the current excess seen at the Large Hadron Collider via the $gg\to H\to\gamma\gamma$ channel 
in the search for a neutral Higgs boson in terms of  both singly (marginally) and doubly (predominantly) charged Higgs bosons predicted by the Higgs Triplet Model.}

\FullConference{Prospects for Charged Higgs Discovery at Colliders\\
		 October 8-11, 2012\\
		 Uppsala University Sweden}

\begin{document}

\section{Introduction}

The presence of a new particle in the Large Hadron Collider (LHC) data is now becoming more and more
certain. In fact, the ATLAS and CMS experimental collaborations 
have announced that they observed a
new boson, with a mass of order 125 GeV \cite{:2012gk,:2012gu},
which is consistent with the Higgs particle $H$ expected in the Standard Model (SM). 
The data used correspond to integrated luminosities of $5.1$
fb$^{-1}$ taken with $\sqrt{s} = 7$ TeV and $5.3$ fb$^{-1}$ taken at $8$
TeV and the search is performed in five decay modes: $H \to \gamma
\gamma$, $ZZ$, $WW$, $\tau^+ \tau^-$ and $b\bar{b}$. An excess of
events with respect to the background is clearly observed in the first two decay modes, whereas 
in the other channels  this is somewhat less clear. 
The bulk of the event rates comes from the $gg$-fusion channel.

Current LHC data are not entirely compatible with the SM Higgs hypothesis though, 
particularly
in the decay channel $H\to \gamma\gamma$. In the Higgs Triplet Model (HTM) there are contributions 
to this decay from loops of doubly charged scalars ($H^{\pm\pm}$) and
singly charged scalars $(H^\pm)$ that are not present in the SM. These additional 
contributions are mediated by the trilinear couplings $H_1 H^{++}H^{--}$ and $H_1 H^{+}H^{-}$, 
where $H_1$ is the lightest CP-even Higgs boson of the HTM. In particular, 
there exists the possibility of constructive interference of the $H^{\pm\pm}$ contribution
with that of the $W^\pm$ contribution (the SM leading term), which enables a 
substantial enhancement of the Branching Ratio (BR) of $H_1\to \gamma\gamma$  in the HTM,
with respect to the SM case.

The magnitude of the contribution of $H^{\pm\pm}$ is essentially determined by
the mass of $H^{\pm\pm}$ ($m_{H^{\pm\pm}}$) and a quartic scalar coupling ($\lambda_1$), with the aforementioned
constructive interference arising for $\lambda_1 < 0$. 
Consequently, the ongoing searches for $H\to \gamma\gamma$ restrict the parameter space of
$[m_{H^{\pm\pm}},\lambda_1]$ more stringently for $\lambda_1 < 0$ than for
the recently-studied case of destructive interference with $\lambda_1 > 0$.
Moreover, if the excess of $\gamma\gamma$ events around a mass of 125 GeV in the LHC searches for  $H\to \gamma\gamma$
is substantiated with larger data samples, and the BR is measured to be somewhat larger than that
for the SM Higgs boson, then such an enhancement could be readily accommodated by the $H_1$ 
state of the HTM if $\lambda_1 <0$. 

Finally, although to a lesser extent, also the contribution from $H^\pm$ mediation can play a significant role 
in the HTM, by producing more marginal, yet still sizeable, increases of the $\gamma\gamma$ rate. Quite apart from the
strength of the $H_1 H^{+}H^{-}$, the main reason for this is clearly
a much reduced coupling to photons with respect the case of $H^{\pm\pm}$ states, owing to a
halved Electro-Magnetic (EM) charge in comparison. 

This write-up collects the salient features of a study carried out in Ref.~\cite{Akeroyd:2012ms} to which we refer the reader for further details. 

\section{The Higgs Triplet Model}

The HTM \cite{Konetschny:1977bn,Mohapatra:1979ia,Magg:1980ut,Schechter:1980gr,Cheng:1980qt}
is a model of neutrino mass generation with a non-minimal Higgs sector.
This scenario predicts several scalar particles, 
including the doubly and singly charged Higgs bosons, for which
direct searches are being  carried out at the LHC \cite{CMS-search,Aad:2012cg}. In a large
part of the parameter space of the HTM the lightest CP-even scalar  has essentially the same couplings 
to the fermions and vector bosons as the Higgs boson of the SM \cite{Dey:2008jm,Akeroyd:2010je,Arhrib:2011uy}.
Therefore the ongoing searches for the SM Higgs boson also apply to $H_1$ of the HTM with very little modification.
An exception is the loop-induced decay $H_1\to \gamma\gamma$ which receives
contributions from a virtual $H^{\pm\pm}$ and $H^{\pm}$ and can have a BR which is very
different from that of the SM Higgs boson. As shown recently in \cite{Arhrib:2011vc},
the ongoing limits on BR($H_1\to \gamma\gamma$) constrain the parameter space of
$[m_{H^{\pm\pm}},\lambda_1]$, where $\lambda_1$ is
a quartic coupling in the scalar potential (see also \cite{Kanemura:2012rs} for a related study). The case of $\lambda_1 > 0$ was studied in
\cite{Arhrib:2011vc}, which leads to destructive interference between the 
combined SM contribution (from $W$ and fermion loops) and the contribution from $H^{\pm\pm}$ (and $H^\pm$).
In this talk we consider the case of $\lambda_1 < 0$, which leads to constructive interference and was not
considered in \cite{Arhrib:2011vc,Kanemura:2012rs}.
The scenario of $\lambda_1 < 0$ is more constrained by the ongoing searches for $H_1\to \gamma\gamma$ than the case of 
$\lambda_1 > 0$. Moreover, the scenario of $\lambda_1 < 0$ can provide enhancements of
 $H_1\to \gamma\gamma$ with smaller $|\lambda_1|$ than the case of $\lambda_1 > 0$.

\section{Numerical Analysis} 

In this section we quantify the magnitude of the charged scalar loops
(due to $H^{\pm\pm}$ and $H^\pm$) on
\begin{equation}
R_{\gamma\gamma}=\frac{{\rm BR}(H \rightarrow \gamma\gamma)^{\rm HTM}} 
{{\rm BR}(H \rightarrow \gamma\gamma)^{\rm SM}}. 
\end{equation}
(Notice that for a SM-like $H_1$ state we have that the $gg$-fusion production rates are practically the same as
for the $H$ state of the SM.)
The case of $\lambda_1>0$ was studied in detail in
\cite{Arhrib:2011uy}. We confirm their results and present new ones
for the case of $\lambda_1<0$.
In our numerical analysis, we have fixed fundamental Lagrangian parameters to obtain
 $m_{H_1}\sim 125$ GeV. In doing so we have taken into account experimental constraints  
from the $\rho$ parameter measurements.
Moreover, our choices  ensure that BR($H^{\pm\pm}\to \ell^\pm\ell^\pm$) is negligible and so
the existing strong limit $m_{H^{\pm\pm}}>400$ GeV does not apply to the HTM due to the dominance of
the decay mode $H^{\pm\pm}\to WW$ (or $H^{\pm\pm}\to H^\pm W^*$ if there is a mass splitting
between $H^{\pm\pm}$ and $H^\pm$). We have further enforced
vacuum stability conditions. In the end, for our setup, $H^{\pm\pm}$ and $H^{\pm}$ are 
essentially degenerate and the couplings $g_{H_1H^{++}H^{--}}$ and $g_{H_1H^+H^-}$
are approximately equal. 

We then treat $\lambda_1$ and $m_{H^{\pm\pm}}$
 as free parameters that
essentially determine the magnitude of
the $H^{\pm\pm}$ and $H^\pm$ contributions to $H_1\to \gamma\gamma$.
We present results for the range:
\begin{eqnarray}
-3 < \lambda_1 < 10,\;\;\;\;\;\;  150~{\rm GeV} < m_{H^{\pm\pm}} < 600~{\rm GeV} \, .
\label{range}
\end{eqnarray}

\begin{figure}[t]
\begin{center}
\includegraphics[origin=c, angle=0, scale=0.5]{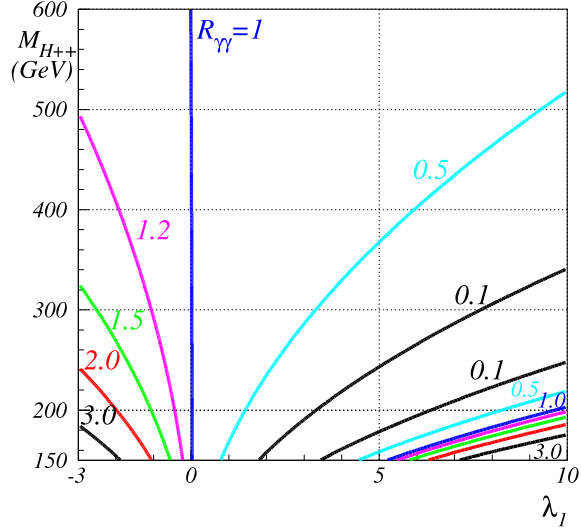}
%\vspace*{-25mm}                                                                                                             
\caption{The ratio $R_{\gamma\gamma}$ in the plane of $[\lambda_1, m_{H^{\pm\pm}}]$
for $150~{\rm GeV} < m_{H^{\pm\pm}} < 600$ GeV, for $m_{H_1}\sim 125$ GeV and
$m_{H^{\pm\pm}}=m_{H^\pm}$.}
\label{rgam1}
\end{center}
\end{figure}

In Fig.~\ref{rgam1}, $R_{\gamma\gamma}$ is plotted in the plane of
$[\lambda_1, m_{H^{\pm\pm}}]$. As  $m_{H^{\pm\pm}}=m_{H^\pm}$  and 
$g_{H_1H^{++}H^{--}}$ $=$ $g_{H_1H^+H^-}$, the
contributions of $H^{\pm\pm}$ and $H^{\pm}$ to the decay rate of $H_1\to \gamma\gamma$
differ only by their EM charge, with the
term induced by $H^{\pm\pm}$ being four times larger at the amplitude level, hence sixteen times 
in the cross section. The range 
 $150~{\rm GeV} < m_{H^{\pm\pm}} < 600$ GeV is plotted here.
For the case of $\lambda_1>0$ one has destructive interference of the $H^{\pm\pm}$ loop
with that of the $W$ loop, leading to a significant suppression of $R_{\gamma\gamma}$ , i.e.,
in the region $0< \lambda_1 < 5$ and $150~{\rm GeV}< m_{H^{\pm\pm}}<300$ GeV there is a large parameter space 
for $R_{\gamma\gamma}<0.5$. For $0< \lambda_1 < 5$ and $400~{\rm GeV}< m_{H^{\pm\pm}}<600$ GeV
(i.e., the mass region which has yet to be probed in the direct searches which 
assume dominance of the decay $H^{\pm\pm}\to \ell^\pm\ell^\pm$)
the suppression is more mild, with $0.5 < R_{\gamma\gamma} < 1$. 
Consequently, for $0< \lambda_1 < 5$ a statistically significant signal for $H_1\to \gamma\gamma$ would
require considerably more integrated luminosity than for the case of the Higgs boson of the SM, and so
detection of $H_1\to \gamma\gamma$ might not be possible in the 8 TeV run of the LHC. However,
the other LHC search channels which make use of the tree-level decays (i.e., $H_1 \to WW, ZZ$, etc.)
would have the same detection prospects as those of the Higgs boson of the SM. 

The case of  $R_{\gamma\gamma}=1$ occurs for $\lambda_1\sim 0$
(i.e., a negligible trilinear coupling $H_1 H^{++}H^{--}$) and also for 
a straight line which joins the points $\lambda_1=5$, $m_{H^{\pm\pm}}=150$ GeV
and $\lambda_1=10$, $m_{H^{\pm\pm}}=200$ GeV. Hence any signal
for $H_1\to \gamma\gamma$ with  $R_{\gamma\gamma}\sim 1$ 
(and assuming $m_{H^{\pm\pm}}<600$ GeV) would
restrict the parameter space of $[\lambda_1, m_{H^{\pm\pm}}]$ to two regions: i)
the region of $\lambda_1\sim 0$, and ii) the region of $\lambda_1 > 5$.
If $H^{\pm\pm}$ is very heavy and out of the discovery reach of the LHC (e.g., $m_{H^{\pm\pm}}>>1$ TeV) then
$R_{\gamma\gamma}\sim 1$ could be accommodated with any positive and sizeable $\lambda_1$.
As emphasised in \cite{Arhrib:2011uy}, 
in the region of $5< \lambda_1 < 10$ and  $m_{H^{\pm\pm}}<200$ GeV, the contribution of the $H^{\pm\pm}$
loop is so large that $R_{\gamma\gamma}>1$ occurs. This region is still compatible
with current LHC data, which excludes $R_{\gamma\gamma}>3.5$ for $m_{H_1}$ around 125 GeV, 
while $R_{\gamma\gamma}> 2$ is excluded for essentially all other choices of
$m_{H_1}$ in the interval $110~{\rm GeV} < m_{H_1}< 150$ GeV.
The excess of $\gamma\gamma$ events at 125 GeV, if assumed to originate from a Higgs boson,
roughly corresponds to $R_{\gamma\gamma}=2.1\pm 0.5$. If $R_{\gamma\gamma}>1$ 
turns out to be preferred by LHC data then one interpretation in the HTM would be
the region of  $5< \lambda_1 < 10$ and  $m_{H^{\pm\pm}}<200$ GeV \cite{Arhrib:2011uy}.

We now discuss the case of $\lambda_1<0$, for which $R_{\gamma\gamma}>1$.
The current sensitivity to $H_1\to \gamma\gamma$ in the LHC searches is
between $1< R_{\gamma\gamma}< 2$ in the mass range $110$ GeV $ < m_{H_1} < 150$ GeV, and 
so the scenario of $\lambda_1<0$ is now being probed by the ongoing searches.
One can see that the current best fit value of $R_{\gamma\gamma}=2.1\pm 0.5$ 
can be accommodated by values of $|\lambda_1|$ which are much smaller than for the case of
$\lambda_1 >0$, e.g., $R_{\gamma\gamma}=2$ can be obtained for $\lambda_1\sim -1$ (or $\lambda_1\sim 6$) 
and $m_{H^{\pm\pm}}=150$ GeV. Importantly, any measured value of $R_{\gamma\gamma}>1$ would
be readily accommodated by the scenario $\lambda_1 < 0$, even for a relatively heavy $H^{\pm\pm}$,
e.g., for $m_{H^{\pm\pm}}>400$ GeV one has
$1.0 < R_{\gamma\gamma}< 1.3$. If either of the
decays $H^{\pm\pm}\to WW$ or $H^{\pm\pm}\to H^\pm W^*$ is dominant (for which there have been
no direct searches) then $m_{H^{\pm\pm}}<400$ GeV is not experimentally excluded, and larger
enhancements of  $R_{\gamma\gamma}$ are possible, 
e.g.,  $R_{\gamma\gamma}= 4.5, 3.1$ and 1.9 for  $\lambda_1=-3, -2$ and $-1$, with $m_{H^{\pm\pm}}=150$ GeV.
Such large enhancements of $R_{\gamma\gamma}$ would require a relatively light $H^{\pm\pm}$
(e.g., $m_{H^{\pm\pm}}<300$ GeV) which decays dominantly to $H^{\pm\pm}\to WW$ and/or $H^{\pm\pm}\to H^\pm W^*$. Simulations of $H^{\pm\pm}\to WW$ were performed in
\cite{Perez:2008ha,Chiang:2012dk}, with good detection prospects for $m_{H^{\pm\pm}}<300$ GeV.
A parton-level study of $H^{\pm\pm}\to H^\pm W^*$ (for the signal only) has been carried out in
\cite{Aoki:2011pz}.

\begin{figure}[t]
\begin{center}
\includegraphics[origin=c, angle=0, scale=0.5]{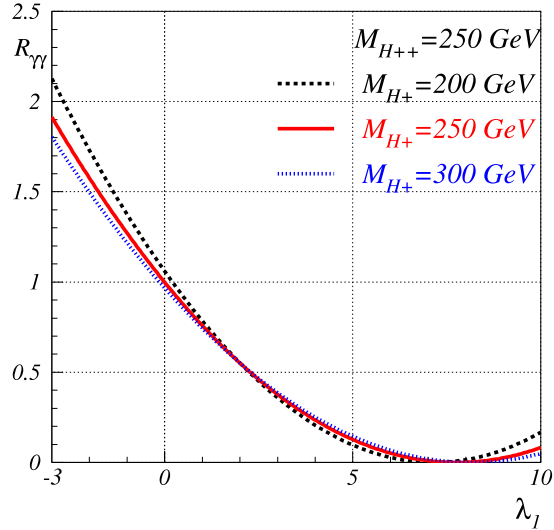}
%\vspace*{-25mm}                                                                                                             
\caption{The ratio $R_{\gamma\gamma}$ as a function of $\lambda_1$ for $m_{H^{\pm\pm}}=250$ GeV,
and $m_{H^\pm}=200$, 250 and 300 GeV, with $m_{H_1}\sim 125$ GeV.}
\label{rgam-hp}
\end{center}
\end{figure}

If one instead allows for $g_{H_1H^{++}H^{--}}$ $\ne$ $g_{H_1H^+H^-}$
 and there is a mass splitting between  $m_{H^{\pm\pm}}$ and
$m_{H^{\pm}}$, the contribution of $H^{\pm\pm}$ to $H_1\to \gamma\gamma$ is not simply 
four times the contribution of $H^\pm$ at the amplitude level. 
In Fig.~\ref{rgam-hp} we plot $R_{\gamma\gamma}$ as a function of $\lambda_1$ , fixing
$m_{H^{\pm\pm}}=250$ GeV and taking $m_{H^\pm}=200$, 250 and 300 GeV. One can see that the case 
of $m_{H^\pm}=200$ GeV and $\lambda_1 < 0$ leads to a value of $R_{\gamma\gamma}$
which is roughly $10\%$ larger than the value for the case of $m_{H^\pm}=m_{H^{\pm\pm}}$. 
For $\lambda_1 > 0$ the magnitude of $g_{H_1 H^+H^-}$ is 
less than that of $g_{H_1 H^{++}H^{--}}$ due to a destructive interference in the decay loop and at around
$\lambda_1=2$ the value of $R_{\gamma\gamma}$ becomes equal to the case of $m_{H^\pm}=m_{H^{\pm\pm}}$. For
$\lambda_1 > 8$ one finds again values of $R_{\gamma\gamma}$ which are slightly larger than for the case
of  $m_{H^\pm}=m_{H^{\pm\pm}}$. For $m_{H^\pm}=300$ GeV the converse dependence 
of $R_{\gamma\gamma}$ on $\lambda_1$ is found.

\section{Conclusions}

In this talk we pointed out that constructive interference of the $H^{\pm\pm}$ contribution with the
$W$ contribution occurs for $\lambda_1<0$ and such a parameter space is consistent with
theoretical constraints on $\lambda_1$ from requiring the stability of the vacuum of the scalar potential. For $m_{H^{\pm\pm}}=400$ GeV, which
is roughly the bound if the decays $H^{\pm\pm}\to \ell^\pm\ell^\pm$  are dominant in the HTM, an enhancement
of up to $\sim 1.3,1.2$ and $1.1$ is possible for $\lambda_1=-3,-2$ and $-1$. Conversely, if either of the
decays $H^{\pm\pm}\to WW$ or $H^{\pm\pm}\to H^\pm W^*$ is dominant (for which there have been
no direct searches so far) then $m_{H^{\pm\pm}}<400$ GeV is not experimentally excluded and larger
enhancements of $\sim 4.5, 3.1$ and $1.9$ are possible for  $\lambda_1=-3,-2$ and $-1$ with $m_{H^{\pm\pm}}=150$ GeV.
Consequently, the parameter space of $\lambda_1 <0$ in the HTM 
is more tightly constrained by the ongoing searches for $H_1 \to \gamma\gamma$ than the case of $\lambda_1 > 0$. 
Importantly, the case of $\lambda_1 <0$ would readily accommodate any signal for $H_1\to \gamma\gamma$ with 
a BR which is higher than that for the SM Higgs boson, 
for  {smaller} values of $|\lambda_1|$ than for the case of $\lambda_1>0$. 
In such a scenario, dedicated searches at the LHC for the decay channels $H^{\pm\pm}\to WW$ or $H^{\pm\pm}\to H^\pm W^*$ with
$m_{H^{\pm\pm}}<400$ GeV would be strongly motivated.

\end{document}